\documentclass[aps,twocolumn]{revtex4}
\usepackage{amsmath}
\usepackage{amssymb}
\usepackage{graphics}
\date{\today}
\begin{document}
\title{Fundamental Limits of the Dispersion of the Two-Photon Absorption Cross-Section}
\author{Javier P\'erez Moreno}\altaffiliation{Present Address: Department of Chemistry, University of Leuven, Celestijnenlaan 200D, B-3001 Leuven, Belgium}\email{Javier.PerezMoreno@fys.kuleuven.be}
\author{Mark G. Kuzyk}\email{kuz@wsu.edu}
\affiliation{Department of Physics, Washington State University \\ Pullman, WA  99164-2814}

\begin{abstract}
We rigorously apply the sum rules to the sum-over-states expression to calculate the fundamental limits of the dispersion of the two-photon absorption cross-section.  A comparison of the theory with the data suggests that the truncated sum rules in the three-level model give a reasonable fundamental limit.  Furthermore, we posit that the two photon absorption cross-section near the limit must have only three dominant states, so by default, the three-level model is appropriate.  This ansatz is supported by a rigorous analytical calculation that the resonant term gets smaller as more states are added.  We also find that the contributions of the non-explicitly resonant terms can not be neglected when analyzing real molecules with many excited states, even near resonance.  However, puzzling as it may be, extrapolating an off-resonant result to resonance using only the resonant term of the three-level model is shown to be consistent with the exact result.  In addition, the off-resonant approximation is shown to scale logarithmically when compared with the full three-level model.  This scaling can be used to simplify the analysis of measurements.  We find that existing molecules are still far from the fundamental limit; so, there is room for improvement.  But, reaching the fundamental limit would require precise control of the energy-level spacing, independently of the transition dipole moments --  a task that does not appear possible using today's synthetic approaches.  So, we present alternative methods that can still lead to substantial improvements which only require the control of the transition moment to the first excited state.  While it is best to normalize measured two photon absorption cross-sections to the fundamental limits when comparing molecules, we show that simply dividing by the square of the number of electrons per molecule yields a good metric for comparison.  
\end{abstract}
\maketitle

\section{Introduction}       

Two-photon absorption is the fundamental process behind most resonant applications such as three-dimensional photolithography,\cite{ch7m:photolitography} photodynamic therapy,\cite{ch7m:photodynamic} high-density storage\cite{ch7m:storage} and optical power limiters.\cite{ch7m:power} All such applications require a large resonant two photon absorption (TPA) cross-section. The development of a deeper understanding of the TPA cross section could be used as a guide to optimizing the nonlinear response of molecules.  Our approach is to use sum rules, which relate transition dipole moments and energies to each other, to learn how the various terms in the sum-over-states expression interfere with each other and how molecules can be designed for constructive interference.

The maximum fundamental limits of the largest diagonal tensor component of the off-resonance first and second hyperpolarizabilities have been calculated using the generalized Thomas-Kuhn sum rules.\cite{Kuzlett,KuzPhysRev,circuits} Extrapolating this off-resonance result of the second hyperpolarizability made it possible to {\em estimate} the maximum values of the resonant two-photon absorption cross-section.\cite{Kuztwopho} Using this process, it is found that most organic molecules show a TPA cross section that falls well below the estimated limit (by a factor of at least $10^{-3/2}$), suggesting that the paradigm for making organic molecules for TPA applications may not be optimized.  This factor appears to be universally observed for single molecules in all nonlinear processes studied (such as second harmonic generation, hyper rayleigh scattering,\cite{kakali} and two photon absorption);\cite{kuzOPN} but can be breached if the molecules are made to interact.\cite{sargent}  In any case, however, the fundamental limit has not been breached, nor is it expected that it would be possible to do so since the sum rules are as basic as the Schr\"odinger equation from which it derives.

In the present work, we apply the sum rules directly to the full dispersion of the TPA cross-section without extrapolation and without neglecting any terms.  As such, these results are exact, in analogy with past calculations of the fundamental limits of the off-resonant hyperpolarizabilities.  We use our approach to analyze a series of molecular measurements which shows that the fundamental limits are somewhat lower than predicted using extrapolation from the off-resonant result; but, the molecules analyzed are still well below the fundamental limits.  More importantly, we show that Two-Photon Fluorescence measurement, used in getting TPA cross-sections, may be inaccurate if all states are not included in the analysis.

\section{Theory}

The TPA cross-section $\delta$ (in units of $cm^{4}s$) is related to the imaginary part of the diagonal component of the second hyperpolarizability, $\gamma_{I}$, through:\cite{Albota}
\begin{equation}
\label{eq:tpacross}
\delta(\omega)=\frac{4\pi^{2}\hbar \omega^{2}}{n^{2}c^{2}} \langle \gamma_{I}^{*} \rangle,
\end{equation}
where $\gamma_{I}= Im(\gamma) \equiv Im(\gamma_{xxxx}(-\omega;\omega,\omega,-\omega))$, $n$ is the refractive index of the bulk material, $\omega$ the frequency of the incident light, $c$ the speed of light, and the brackets indicate the average over all-possible orientations of the molecule. For a one-dimensional molecule $\langle \gamma \rangle = \gamma/5$ and for a spherical molecule $\langle \gamma \rangle = \gamma$. The effects of the local fields are included in the dressed hyperpolarizability, $\gamma_{I}^{*}$.

Two photon absorption is inherently a resonant process because it quantifies the strength of absorption of two photons - each with energy $\hbar \omega$ - from the ground state of energy $E_0$ to the second excited state with energy $E_2$, where $E_{20} \equiv E_2 - E_0 = 2\hbar \omega$.  Consequently, the exact expression for the diagonal component of the second hyperpolarizability given by the sum-over-states expression\cite{OrrWard} is often approximated by the explicitly resonant term in the sum, $\gamma_{res}$:
\begin{equation}
\label{resonant}
\gamma \approx \gamma_{res} \equiv \frac{ \mu_{01}^{2}\mu_{12}^{2}}{(E_{10}-\hbar\omega-i\Gamma_{10})^{2}(E_{20}-2\hbar\omega-i\Gamma_{20})},
\end{equation}
where $E_{n0}$ is the transition energy between the excited state $n$ and the ground state, $\mu_{nm}$ is the transition dipole moment between states $n$ and $m$, and $\Gamma_{n0}$ is the damping factor (inverse radiative lifetime) between the state $n$ and the ground state.  When the approximation $\gamma \approx \gamma_{res}$ is made, the contributions of all the non-explicitly resonant terms to $\gamma$ are ignored. For such an approximation to be valid, $\gamma_{res}$ should dominate the response in the domain of interest for TPA processes. 

Two different regimes can be distinguished in TPA processes: the single resonance regime and the double resonance regime.  In both cases $E_{20} \approx 2 \hbar \omega$; but, in the single resonance regime, the photon energy $\hbar \omega$ does not match the transition energy $E_{10}$ (i.e. $\left| E_{10} - \hbar \omega \right| \gg \Gamma_{10} $ )  while in the double resonance regime $E_{10} \approx \hbar \omega$.  For $\gamma_{res}$ to dominate the second hyperpolarizability $\gamma$, it is necessary to work in the single-resonance regime to ensure that $\gamma_{res}$ is the only ``resonant'' term in the sum-over-states expression. 

Studies of the first and second hyperpolarizabilities in the off-resonance regime clearly show that individual terms in the sum-over-states (SOS) expression cannot be independently adjusted since the terms are related to each other through the sum rules.\cite{Kuzlett,KuzPhysRev} In fact, in the SOS expression for $\gamma$ there are additional two-photon resonant terms similar to Eq. \ref{resonant} that contribute to $\gamma$.  Even if certain terms appear to dominate the response, all the contributions must be included since the contribution of just one term might not be representative of the full result - especially if there are cancellations between large terms.  Therefore, our calculations of the maximum limit of the TPA cross-section includes all contributions to $\gamma$.  We also evaluate the relevance of the different terms in the SOS expression with the aim of determining whether or not Eq. \ref{resonant} is in general a good approximation to $\gamma_{I}$; an approximation that is used in determining the TPA cross-section  (and $\mu_{12}$) with TPA  fluorescence measurements.

The exact sum-over-states expression for $\gamma_{xxxx}(-\omega;\omega,\omega,-\omega)$\cite{OrrWard} can be split into two types of contributions as follows:
\begin{equation}
\label{split1}
\gamma_{xxxx}(-\omega;\omega,\omega,-\omega) \equiv S_{1}+S_{2}+S_{3}+S_{4}+T_{1}+T_{2}+T_{3}+T_{4},
\end{equation}
where
\begin{eqnarray}
\label{eq:S1}
S_{1} &=& (\hbar)^{-3} I_{123} {\sum_{lmn}}' \frac{\mu_{gl} \bar{\mu}_{lm}\bar{\mu}_{mn}\mu_{ng}}{(\Omega_{lg}-\omega)(\Omega_{mg}-2 \omega)(\Omega_{ng}-\omega)}, \\
\label{eq:S2}
S_{2} &=& (\hbar)^{-3} I_{123} {\sum_{lmn}}' \frac{\mu_{gl}\Bar{\mu}_{lm}\bar{\mu}_{mn}\mu_{ng}}{(\Omega^{*}_{lg}-\omega)(\Omega_{mg}-2 \omega)(\Omega_{ng}-\omega)}, \\ 
\label{eq:S3}
S_{3} &=& (\hbar)^{-3} I_{123} {\sum_{lmn}}' \frac{\mu_{gl}\Bar{\mu}_{lm}\bar{\mu}_{mn}\mu_{ng}}{(\Omega^{*}_{lg}+\omega)(\Omega^{*}_{mg}+2 \omega)(\Omega_{ng}+\omega)}, \\
\label{eq:S4}
S_{4} &=& (\hbar)^{-3} I_{123} {\sum_{lmn}}' \frac{\mu_{gl}\Bar{\mu}_{lm}\bar{\mu}_{mn}\mu_{ng}}{(\Omega^{*}_{lg}+\omega)(\Omega^{*}_{mg}+2 \omega)(\Omega^{*}_{ng}+\omega)}, \\
\label{eq:T1}
T_{1} &=& -(\hbar)^{-3} I_{123}{\sum_{mn}}'\frac{\mu_{gm}\mu_{mg}\mu_{gn}\mu_{ng}}{(\Omega_{mg}-\omega)(\Omega_{mg}+\omega)(\Omega_{ng}-\omega)}, \\
\label{eq:T2}
T_{2} &=& -(\hbar)^{-3} I_{123}{\sum_{mn}}'\frac{\mu_{gm}\mu_{mg}\mu_{gn}\mu_{ng}}{(\Omega_{mg}+\omega)(\Omega^{*}_{ng}+\omega)(\Omega_{ng}-\omega)}, \\
\label{eq:T3}
T_{3} &=& -(\hbar)^{-3} I_{123}{\sum_{mn}}'\frac{\mu_{gm}\mu_{mg}\mu_{gn}\mu_{ng}}{(\Omega^{*}_{mg}+\omega)(\Omega^{*}_{mg}-\omega)(\Omega^{*}_{ng}+\omega)}, \\
\label{eq:T4}
T_{4} &=& -(\hbar)^{-3} I_{123}{\sum_{mn}}'\frac{\mu_{gm}\mu_{mg}\mu_{gn}\mu_{ng}}{(\Omega^{*}_{mg}-\omega)(\Omega_{ng}-\omega)(\Omega^{*}_{ng}+\omega)}.
\end{eqnarray}
Here, $I_{123}$ denotes the average over all distinct permutations of $\omega_{1}$, $\omega_{2}$ and $\omega_{3}$, where  $\omega_{1}=\omega$, $\omega_{2}=\omega$ and $\omega_{3}=-\omega$. The prime in the sum indicates that the sum is over the excited states, $\Omega_{n0}$ is a complex quantity defined as:
\begin{equation}
\hbar \Omega_{n0}=E_{n0}-i\Gamma_{n0},
\end{equation}
and:
\begin{equation}
\Bar{\mu}_{lm}=\left\{ \begin{array}{cc}
        \mu_{lm} & \mbox{for $l \neq m$} \\
        \mu_{mm}-\mu_{00} & \mbox{for $l=m$}.
        \end{array}
    \right.
\end{equation}
Note that the explicitly two-photon single-resonant terms (when $E_{20}=2 \hbar \omega$) are all contained in $S_{1}$ and $S_{2}$ (Eqs. \ref{eq:S1} and \ref{eq:S2}). $T_{1}$ and $T_{2}$ (Eqs. \ref{eq:T1} and \ref{eq:T2}) contain terms that will be resonant in the double-resonance regime (when $E_{10}=\hbar \omega$ and $E_{20}= 2 \hbar \omega$).

It is important to keep sight of the approximations inherent in the SOS expression, which is at the heart of all our calculations.  It assumes that the light emitted by a molecule is electric dipole in nature, so magnetic transitions as well as higher-order electric terms are ignored.  In addition, since the SOS expression is derived from perturbation theory, the energy of interaction between the light and the molecule must be small compared with the binding energy.  As such, the light can not excite the molecule into a real high-lying state.  Finally, we focus on purely electronic transitions, so vibronics are not considered.  However, it has been shown that neglect of nuclear motion has only a small affect on the limits of the calculated nonlinear-optical response.\cite{kakali}

We consider a general three-level model and truncate the Thomas-Kuhn sum rules to the first three energy levels to obtain relationships between the different parameters that determine $\gamma$. The procedure is similar to the one followed to obtain the maximum value of the off-resonance first and second hyperpolarizabilities.\cite{Kuzlett,KuzPhysRev} Unlike this past work, the present work includes the effects of dispersion, so the widths of each state are included by using an imaginary part of the frequency, as shown in Equations \ref{eq:S1} through \ref{eq:T4}.  Hence, the expressions are mathematically more algebraically messy.  We therefore use {\sl MATHEMATICA$^{\textregistered}$} to evaluate the results.

The full set of self-consistent three-level truncated Thomas-Kuhn sum rules yield relationships between transition dipole moments and energy differences as follows:
\begin{eqnarray}
\label{sr1}
E_{10}\mu_{10}^{2}+E_{20}\mu_{20}^{2} &=& \frac{(\hbar e)^{2}N}{2m}, \\
\label{sr2}
(2E_{20}-E_{10})\mu_{20}\mu_{21} + E_{10} \mu_{10} \Delta \mu_{10} &=& 0, \\
\label{sr3}
-E_{10} \mu_{10}^{2} + (E_{20}-E_{10}) \mu_{21}^{2} &=& \frac{(\hbar e)^{2}N}{2m}, \\
\label{sr4}
(2E_{10}-E_{20}) \mu_{10} \mu_{21} + E_{20} \mu_{20} \Delta \mu_{20} &=& 0,
\end{eqnarray}
where $\Delta \mu_{mn} = \mu_{mm} - \mu_{nn}$, $e$ is the charge of an electron, $N$ is the number of electrons in the molecule and $m$ is the mass of an electron. 

A direct consequence of the first sum rule given by Eq. \ref{sr1}, is that the values of $|\mu_{10}|$ are constrained:\cite{Kuzlett}
\begin{equation}
\mu_{10}^{2} \leq \frac{(\hbar e)^{2}N}{2mE_{10}} \equiv |\mu_{10}^{MAX}|^{2}.
\end{equation}
Therefore, as in the off-resonance analysis, we define the dimensionless quantity, $X$:
\begin{equation}
X=\frac{|\mu_{10}|}{|\mu_{10}^{MAX}|} \leq 1,
\end{equation}
where $-1 \leq X \leq 1$.  We also define the dimensionless quantity $E$ as:
\begin{equation}
E=\frac{E_{10}}{E_{20}},
\end{equation}
with $0 \leq E \leq 1$.  We note that this explicitly assumes that $E_{10} \leq E_{20}$ and that $E_{20}$ is the two-photon state.  These criteria are obeyed for all of the molecules analyzed in this paper.  Note that systems such as the longer polyenes have lower-energy two-photon states.  As such, a separate calculation in the spirit of the one that follows would need to be developed.  While we do not do such a calculation in this paper since it is not relevant to our studies, interesting results are possible.  Secondly, we note that if the first two excited states in an asymmetric molecule become degenerate, then the next excited state will need to be considered and the energy ratio redefined, i.e. $E \rightarrow E_{10} / E_{30}$.

Our objective is to express $\gamma_{I}$ in terms of $X$ and $E$ and determine the fundamental limits of the TPA cross-section, $\delta(\omega)$, by finding the values of $X$ and $E$ that maximize $\delta$.  It is important to emphasize that the calculation of the fundamental limits makes only one assumption: the forces between the charges are described by a potential energy that depends exclusively on the coordinates of the electrons and the nuclei.  The tools that we use to calculate the fundamental limits, the generalized Thomas-Kuhn sum rules, are fundamental identities derived from the Sch\"odinger equation, and are applicable to all electronic excitations, regardless of the particular details of the molecule. However, we approximate the molecule as a three-level system to model the response and truncate the sum rules to three states, an approach that has been successful at determining the fundamental limits in the off-resonance regime.\cite{Kuzlett,KuzPhysRev} In fact, recent research suggests that molecules with more excited states have a smaller first hyperpolarizability than the three-level system.\cite{kakali} As such, the limits we calculate are truly an upper limit.

\subsection{Contribution of the explicitly resonant two-photon terms}

We first evaluate the contributions of the terms $Im(S_{1})$ and $Im(S_{2})$ to $\gamma_{I}$, since they contain all the two-photon resonant contributions (when $E_{20}=2 \hbar \omega$). Using the three-level-truncated Thomas-Kuhn sum rules, we are able to parameterize $Im(S_{1})$ and $Im(S_{2})$ in terms of $E_{10}$, $E_{20}$, $X$, $\Gamma_{10}$, and $\Gamma_{20}$. Since these expressions are constrained by the three-level Thomas-Kuhn sum rules, they are labelled $S_i^{TK3}$.  We assume at all times that the two-photon resonance condition is obeyed, that is $E_{20}=2 \hbar \omega$; and we vary the other parameters.

We find that both $Im(S_{1}^{TK3})$ and $Im(S_{2}^{TK3})$ vanish when $X=0$ and reach their maximum value when $X=1$. Also, as $\Gamma_{10}$ and $\Gamma_{20}$ increase as a fraction of $E_{10}$, the values of $Im(S_{1}^{TK3})$ and $Im(S_{2}^{TK3})$ approach zero. For our calculations we consider only one electron (N=1) and we use typical values of the parameters for organic molecules:
\begin{eqnarray}
\label{mutypical}
X &=& 0.5, \\
\label{gammatypical}
\Gamma_{10} &=& \Gamma_{20} = 0.1 eV, \\
\label{etypical}
\mbox{ and  } E_{10} &=& 2eV .
\end{eqnarray} 

\begin{figure}
\includegraphics{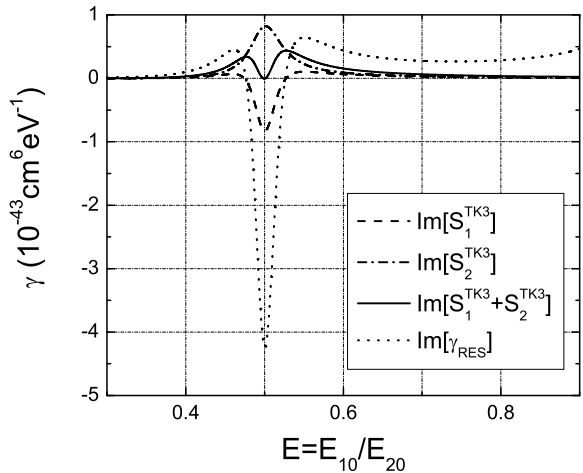}
\caption{$Im(S_{1}^{TK3})$, $Im(S_{2}^{TK3})$, $Im(S_{1}+S_{2})$ and $Im(\gamma_{res})$ as a function of the energy ratio, $E$, with $X=0.5$, $\Gamma_{10} = \Gamma_{20} = 0.1 eV$, $E_{10}=2eV$, and $N=1$.\label{fig:UNO}}
\end{figure}

Fig. \ref{fig:UNO} shows a plot of $Im(S_{1}^{TK3})$ and $Im(S_{2}^{TK3})$ individually as a function of the energy ratio $E=E_{10}/E_{20}$. Also plotted is the behavior of their sum, $Im(S_{1}^{TK3})+Im(S_{2}^{TK3})$, and $Im(\gamma_{res})$ (Eq. \ref{resonant}). 
Both $Im(S_{1}^{TK3})$ and $Im(S_{2}^{TK3})$ are peaked at $E=1/2$, and fall quickly to zero as $E \rightarrow 1$ and $E \rightarrow 0$.  As we later discuss, this is the general trend for all the terms that contribute to $\gamma_{I}$, resulting in an imaginary third-order susceptibility that peaks around $E=1/2$. As such, to maximize the TPA cross-section, the two lowest excited state energies of a molecule must be equally spaced to allow for the double-resonance condition ($E_{20}= 2 E_{10} = 2 \hbar \omega$).\cite{kuzres}

While individually $Im(S_{1}^{TK3})$ and $Im(S_{2}^{TK3})$ are peaked functions at $E=1/2$, they have opposite signs, so these two terms partially cancel. More importantly, although $Im(S_{1}^{TK3}+S_{2}^{TK3})$ contains the term $Im(\gamma_{res})$, it is obvious from the spectrum predicted by $Im(S_{1}^{TK3}+S_{2}^{TK3})$ that it is quantitatively and qualitatively different from $Im(\gamma_{res})$.  Furthermore, $Im(\gamma_{res})$ diverges as $E \rightarrow 1$, while $Im(S_{1}^{TK3})$ and $Im(S_{2}^{TK3})$ approach zero. When all terms are included, the divergences from individual terms in $Im(S_{1}^{TK3})$ and $Im(S_{2}^{TK3})$ cancel in the same way as it does in the off-resonance calculations.\cite{Kuzlett,KuzPhysRev}

\subsection{Full three-level calculation of $\gamma_{I}$}

\begin{figure}
\includegraphics{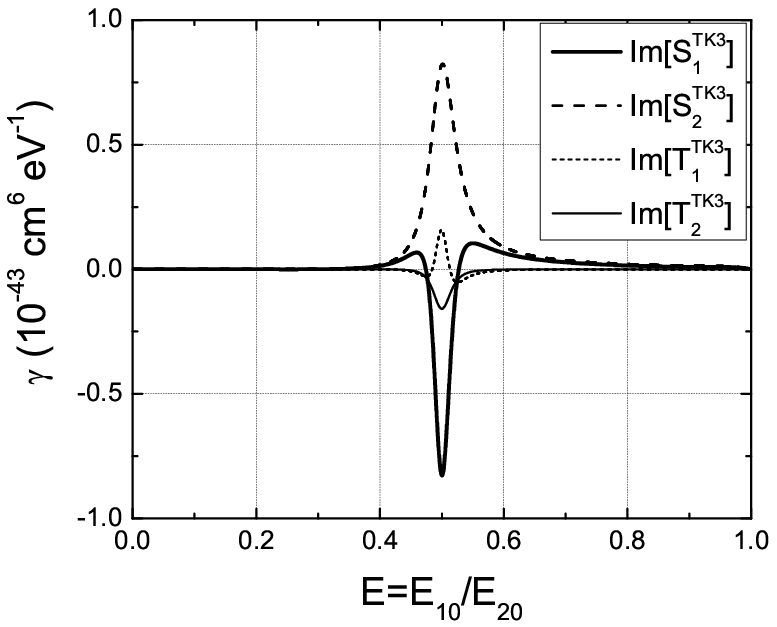}
\caption{$Im(S_{1}^{TK3})$, $Im(S_{2}^{TK3})$, $Im(T_{1}^{TK3})$ and $Im(T_{2}^{TK3})$ as a function of the energy ratio, $E$, with $X=0.5$, $\Gamma_{10} = \Gamma_{20} = 0.1 eV$, $E_{10}=2eV$, and $N=1$.\label{fig:DOSA}}
\end{figure}

\begin{figure}
\includegraphics{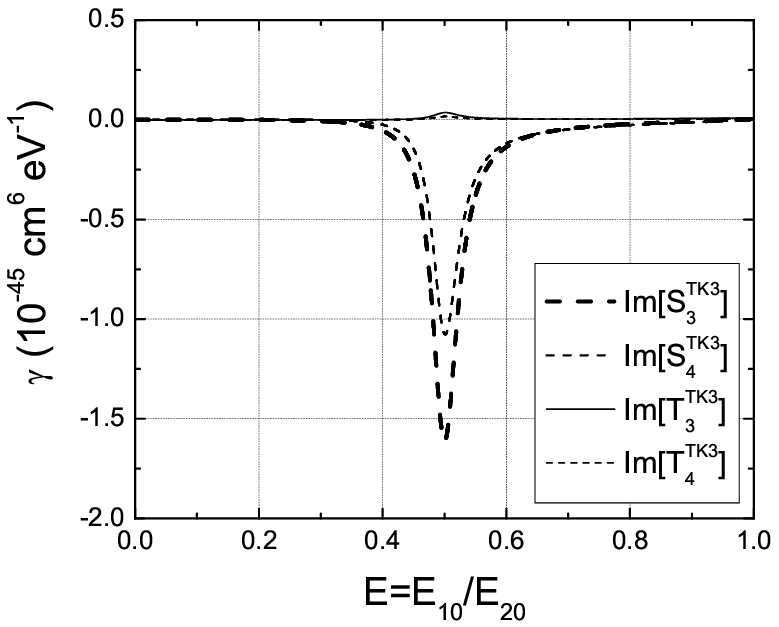}
\caption{$Im(S_{3}^{TK3})$, $Im(S_{4}^{TK3})$, $Im(T_{3}^{TK3})$ and $Im(T_{4}^{TK3})$ as a function of the energy ratio, $E$, with $X=0.5$, $\Gamma_{10} = \Gamma_{20} = 0.1 eV$, $E_{10}=2eV$, and $N=1$.\label{fig:DOSB}}
\end{figure}

The analysis of the sum $Im(S_{1}^{TK3})+Im(S_{2}^{TK3})$ shows that to get a substantial TPA cross-section the system must be close to the double-resonant condition (when $E_{20}= 2E_{10}= 2 \hbar \omega$). In this regime, contributions of terms such as $Im(T_{1}^{TK3})$ and $Im(T_{2}^{TK3})$ can play an essential role, since they explicitly contain one-photon resonant terms ($E_{10}= \hbar \omega$). Therefore, we now study the behavior, individually, of all the terms that contribute to the second hyperpolarizability as constrained by the three-level truncated sum rules, $\gamma_{I}^{TK3}$. We plot each term (defined by Eqs. \ref{eq:S1}-\ref{eq:T4}) as a function of $E$ in Figs. \ref{fig:DOSA} and \ref{fig:DOSB}. For simplicity we have used the values defined in Eqs. \ref{mutypical}, \ref{gammatypical} and \ref{etypical} since these are the typical values for organic molecules that absorb light in the visible. 

Before proceeding, we must take a small detour to deal with the fact that on double resonance, we must consider the other excited states that may contribute to the SOS expression for $\gamma_I$.  This is certainly a crucial issue when modelling real molecules, which have many excited states beyond the first three.  However, we must also recall that our analysis focuses on understanding the fundamental limits.  Thus, our calculations are based on the ansatz that the hyperpolarizability, of any order, will be maximized for any system when all of the transitions are concentrated into three states.\cite{kuzykRes}  As such, when we are dealing with {\em ideal molecules} that are near the fundamental limit, the problems of higher lying states contributing to $\gamma_I$ is moot.  The next section addresses this issue for the explicitly-resonant term.

While the contributions of $Im(S_{3}^{TK3})$, $Im(S_{4}^{TK3})$, $Im(T_{3}^{TK3})$ and $Im(T_{4}^{TK3})$ are negligible in comparison with $Im(S_{1}^{TK3})$ or $Im(S_{2}^{TK3})$, it is obvious from the plots that the contributions of $Im(T_{1}^{TK3})$ and $Im(T_{2}^{TK3})$ are comparable to the contributions of $Im(S_{1}^{TK3})$ and $Im(S_{2}^{TK3})$. This is so because $T_{1}^{TK3}$ and $T_{2}^{TK3}$ contain one-photon resonant contributions.  Since all the terms have different signs around $E \approx 1/2$ it is impossible to estimate the value of $\gamma_{I}^{TK3}$ by just looking at one individual term.

\begin{figure}
\includegraphics{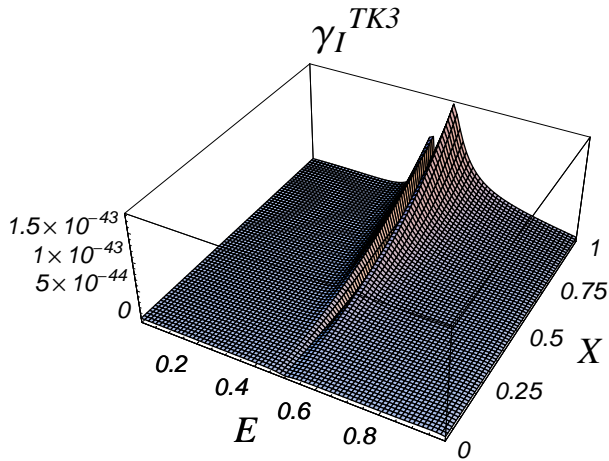} 
\caption{$\gamma_{I}^{TK3}$ as a function of $E$ and $X$, with $\Gamma_{10}=\Gamma_{20}=0.1eV$, $E_{10}=2eV$, and $N=1$.\label{fig:THREE}}
\end{figure}

Fig. \ref{fig:THREE} shows a three-dimensional plot of $\gamma_{I}^{TK3}$ as a function of $E$ and $X$ with $\Gamma_{10} = \Gamma_{20} = 0.1eV$, and $E_{10} = 2 eV$. $\gamma_{I}^{TK3}$ has been plotted for all possible values of $E$ and $X$ to show that indeed, $\gamma_{I}^{TK3}$ is zero when $X=0$ and reaches its maximum value when $X=1$.  In terms of the energy ratios, the function is sharply peaked around $E \approx 1/2$.  $\gamma_{I}^{TK3}$ also gets smaller and approaches zero as the values of $\Gamma_{10}$ and $\Gamma_{20}$ increase relative to $E_{10}$.  In order to obtain bigger values of $\gamma_{I}^{TK3}$, the ratio between $\Gamma_{10}$ (and $\Gamma_{20}$) and $E_{10}$ has to be minimized.

\begin{figure}
\includegraphics{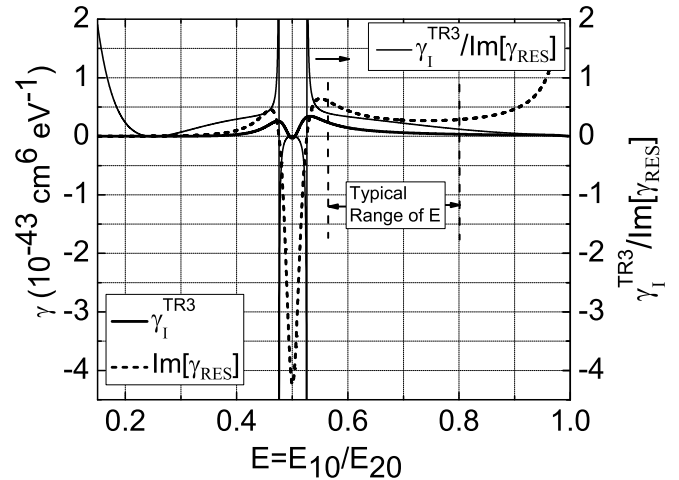} 
\caption{A comparison between $\gamma_{I}^{TK3}$ and $Im(\gamma_{res})$ as a function of $E$, with parameters that are typical for organic molecules: $X=0.5$, $E_{10}=2eV $, $\Gamma_{10} = \Gamma_{20} = 0.1 eV$, and $N=1$.\label{fig:FIVE}}
\end{figure}

In order to compare $\gamma_{I}^{TK3}$ and $Im(\gamma_{res})$ we plot the two functions together in Fig. \ref{fig:FIVE} along with their ratio. $\gamma_{I}^{TK3}$ and $Im(\gamma_{res}$) are both
calculated as a function of $E$, with $X=0.5$, $\Gamma_{10}=\Gamma_{20}=0.1eV$ and $E_{10}=2.0eV$ (typical values).  The typical range of $E$ for many organic molecules is also shown. From the plot, it is obvious that the approximation that $Im(\gamma_{res})$ dominates $\gamma_{I}^{TK3}$ is not a good one for most values of $E$; and, in the range of typical $E$, $\gamma_I$ is smaller than $Im(\gamma_{res})$ by as much as a factor of 3. Clearly, there are big differences between the two at various values of $E$.  For example, while $\gamma_{I}^{TK3}$ remains well behaved and goes to zero at $E \rightarrow 1$, $Im(\gamma_{res})$ diverges. Furthermore, while $\gamma_{I}^{TK3}=0$ at $E=1/2$, $Im(\gamma_{res}) \approx - 4 \times 10 ^{-43} cm^6 eV^{-1}$. We conclude that $\gamma_{I}^{TK3}$ can not be approximated on or near double resonance by $Im(\gamma_{res})$.  For the typical values of $E$ that are found in organic molecules, the two are - on average - different by a factor of 3.

In conclusion, for a given value of $E_{10}$, in order to maximize $\gamma_{I}^{TK3}$ we must maximize the value of the transition moment to the first excited state ($X = 1$), minimize the damping factors (i.e. the inverse radiative lifetime) as a fraction of $E_{10}$ and operate near double resonance, where the imaginary part of second hyperpolarizability is a maximum (when $E_{10} \approx E_{20}/2$). The influence of the values of the inverse radiative lifetimes is evaluated in Table \ref{table1}, where we list the maximum values of $\gamma_{I}^{TK3}$ together with the value of $E$ that maximizes $\gamma_{I}^{TK3}$ for different ratios between the inverse radiative lifetimes and $E_{10}$.

\begin{table}
\caption{Maximum value of $\gamma_{I}^{TK3}$ as a function of the inverse radiative lifetime. $E^{max}$ is the value of $E$ that maximizes $\gamma_{I}^{TK3}$. We have used typical values of the parameters for organic chromophores of: $X=0.5$ and $E_{10}=2eV$.}
\label{table1}
\begin{eqnarray*}
\begin{array}{|c|c|c|} \hline
& &  \\ 
\Gamma_{10}=\Gamma_{20} & \gamma_{I}^{max} & E^{max} \\
& & \\
(eV) & (10^{-44} cm^{6} \cdot eV^{-1}) & \\ 
& &  \\ \hline \hline
0.02 & 387 & 0.49  \\ \hline
0.04 & 51.9 & 0.51  \\ \hline
0.06 & 15.6 & 0.52  \\ \hline
0.08 & 6.65 & 0.53  \\ \hline
0.10 & 3.42 & 0.53 \\ \hline
0.12 & 1.98 & 0.54  \\ \hline
0.14 & 1.25 & 0.55  \\ \hline 
0.16 & 0.829 & 0.56 \\ \hline
0.18 & 0.576 & 0.57 \\ \hline
0.20 & 0.414 & 0.58 \\ \hline \hline
\end{array}
\end{eqnarray*}
\end{table}

\section{Maximum value of $\gamma_{res}$ for molecules with three or more energy levels}

As discussed above, it could be argued that the limits obtained using a three-level model do not necessarily hold for real molecules with more than three energy levels.  That is, it is in principle possible that when a molecule has more levels, the energy distributions allowed by the sum-rules result in a higher limit than the values of $\gamma_{I}^{max}$ listed in table \ref{table1}. This seems unlikely, since, as we will see in the next section, all the measured values of $\delta(\omega)$ are much lower than the maximum limit values allowed by the sum-rules, indicating that real molecules, with more than three available levels, do much more poorly in terms of TPA cross-sections than the idealized three-level model.

We argue that the three-level model gives the best possible value of $\gamma_{I}^{max}$. In other words, as the oscillator strengths are spread out over more excited levels, the value of $\gamma_{I}^{max}$ decreases.  Although this can not be proven rigorously for the case of $\gamma_{I}^{TK \infty}$ using all the SOS contributing terms, it can be shown that if the values of $E_{10}$ and $E_{20}$ are known, the explicitly resonant term, $\gamma_{res}$ is a bounded quantity for {\bf any number of levels} available and that, indeed, the best performance of $\gamma_{res}$ is achieved when only three states are available.  

To prove our assertion we start with the expression for the explicitly resonant term, Eq. \ref{resonant}, which, in the TPA regime $E_{20}=2 \hbar \omega$, can be written as: 
\begin{equation}
\label{eq:resonant2}
\gamma \approx \gamma_{res} \equiv \frac{ \mu_{10}^{2}\mu_{12}^{2}}{(E_{10}-\frac{E_{20}}{2}-i\Gamma_{10})^{2}(-i\Gamma_{20})}.
\end{equation}
Since we assume that $E_{10}$ and $E_{20}$ have fixed values, we optimize $\gamma_{res}$ using $\mu_{10}^{2}$ and $\mu_{12}^{2}$ as adjustable parameters. 

Now we apply the Thomas-Khun sum-rules to a general system with $N'$-levels. The sum-rules can be written as:
\begin{equation}
\sum_{i}^{N'}(2E_{i}-E_{k}-E_{l})\mu_{ki}\mu_{il} = \frac{e^2 \hbar^{2}N}{m}\delta_{kl},
\end{equation}
where $\delta_{kl}$ is the Kronecker delta. We generate a sum-rule equation by choosing one pair of values of $(k,l)$.  They can each be any integer between $0$ and $N'$.

First we consider the choice of parameters $(k,l)=(0,0)$, which yields,
\begin{equation}
\label{eq:firstsumrule}
E_{10}\mu_{10}^{2}+E_{20}\mu_{20}^{2}+E_{30}\mu_{30}^{2}+\cdots+E_{N'0}\mu_{N'0}^{2}=\frac{(\hbar e)^{2}N}{2m}.
\end{equation}
Since all the terms on the left side of the sum are positive, it is obvious that $\mu_{10}^{2}$ and $\mu_{20}^{2}$ will be maximized when:
\begin{equation}
\label{eq:firstmaxcond}
E_{30}\mu_{30}^{2}+E_{40}\mu_{40}^{2}+\cdots+E_{N'}\mu_{N'0}^{2}=0,
\end{equation}
that is, when only states $0$, $1$ and $2$ are present.

Now we consider the combination $(k,l)=(1,1)$, which generates the following sum-rule:
\begin{equation}
-E_{10}\mu_{10}^{2}+E_{21}\mu_{10}^{2}+E_{31}\mu_{31}^{2}+ \cdots + E_{N'1}\mu_{N'1}^{2}=\frac{(\hbar e)^{2}N}{2m}, 
\end{equation}
which can be rewritten as:
\begin{equation}
\label{eq:secondsumrule}
E_{21}\mu_{21}^{2}+E_{31}\mu_{31}^{2}+ \cdots + E_{N'1}\mu_{N'1}^{2} = \left( \frac{(\hbar e)^{2}N}{2m}+E_{10}\mu_{10}^{2} \right).
\end{equation}
All the terms on the left side sum on Eq. \ref{eq:secondsumrule} are positive which implies that $\mu_{21}^{2}$ will be maximum when:
\begin{equation}
\label{eq:secondmaxcond}
E_{31}\mu_{31}^{2}+ \cdots + E_{N'1}\mu_{N'1}^{2} = 0,
\end{equation}
which again would imply that only the states $0$, $1$ and $2$ contribute.  Therefore, if more states contribute and the result is an enhancement of the response, it would necessarily be due to the contribution of terms exclusive of the $\gamma_{res}$ term since we have proven that $\gamma_{res}$ gets smaller as the number of levels increases.

To summarize, we have found that the non-explicitly-non-resonant terms can not be neglected in a strict three-level system, and that if adding more states results in an increase of the TPA response, it will be due to the contribution of the non-explicitly-resonant terms in the SOS expression.   

\section{Applications}

In this section we apply the theoretical results to evaluate the literature values of the measured TPA performance of organic molecules. We do so by comparing the measured value of the TPA cross-section of a real molecule with the maximum value of the TPA cross-section allowed by the sum-rules.  The effective number of electrons contributing to the nonlinear response is calculated by geometrically weighting the number of electrons in each conjugated path of the molecule,\cite{Kuztwopho}
\begin{equation}\label{Neff-count-electrons}
N_{eff} = \sqrt{\sum_{i} N_i^2} ,
\end{equation}
where $N_i$ is the number of electrons in the $i^{th}$ conjugated part of the molecule.  We note that our method for counting electrons is most appropriate for excitations that are typical in conjugated molecules.  For other systems or other types of excitations, an appropriate method for counting electrons would need to be used.

\begin{figure}
\includegraphics{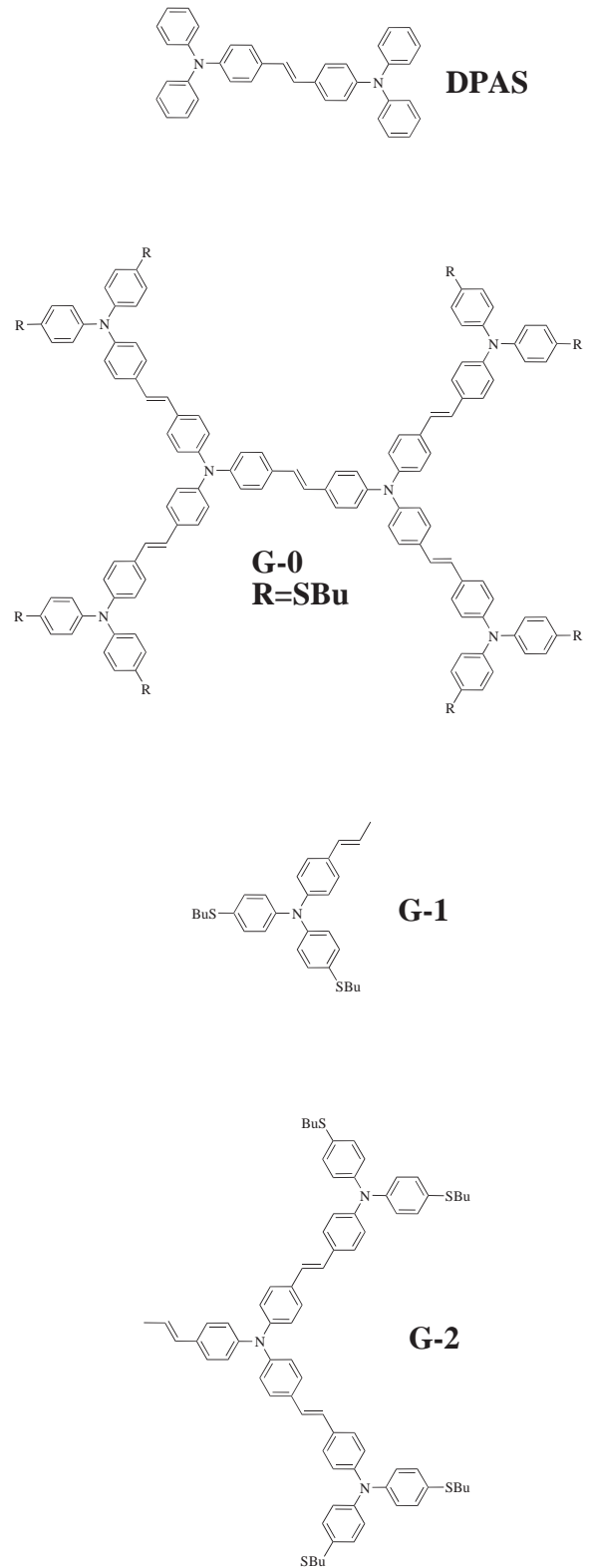} 
\caption{Molecules measured by Dhrovizhev and coworkers.  From top to bottom, the molecules are {\bf 20} and {\bf 21}.  The structures labelled G-1 and G-2 are the $R$-groups, that when added to molecule {\bf 21}, makes the dendrimers {\bf 22} and {\bf 23}.  The effective number of electrons are 18.4, 35.6, 55.8, and 82.7, respectively\cite{Drobizhev}\label{fig:Dhrovizhev}}
\end{figure}

\begin{figure}
\includegraphics{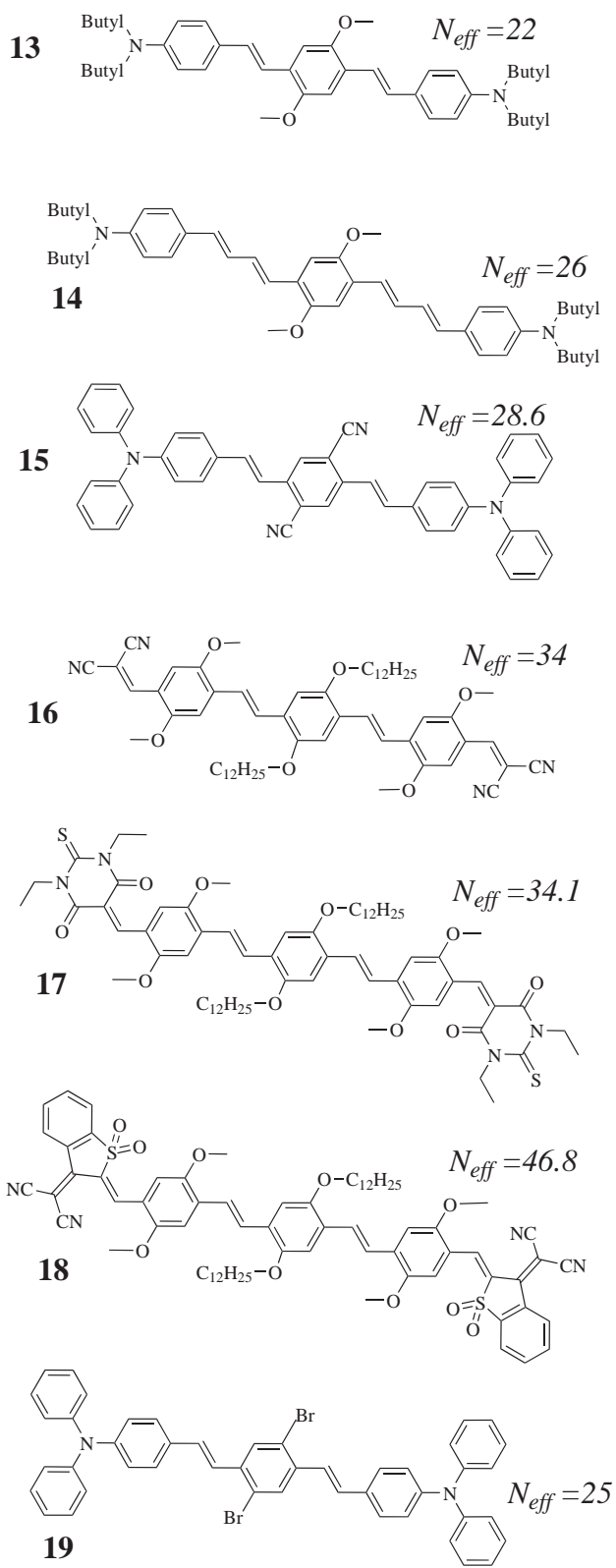} 
\caption{Molecules measured by Albota and coworkers.\cite{Albota}\label{fig:albota}}
\end{figure}

\begin{figure}
\includegraphics{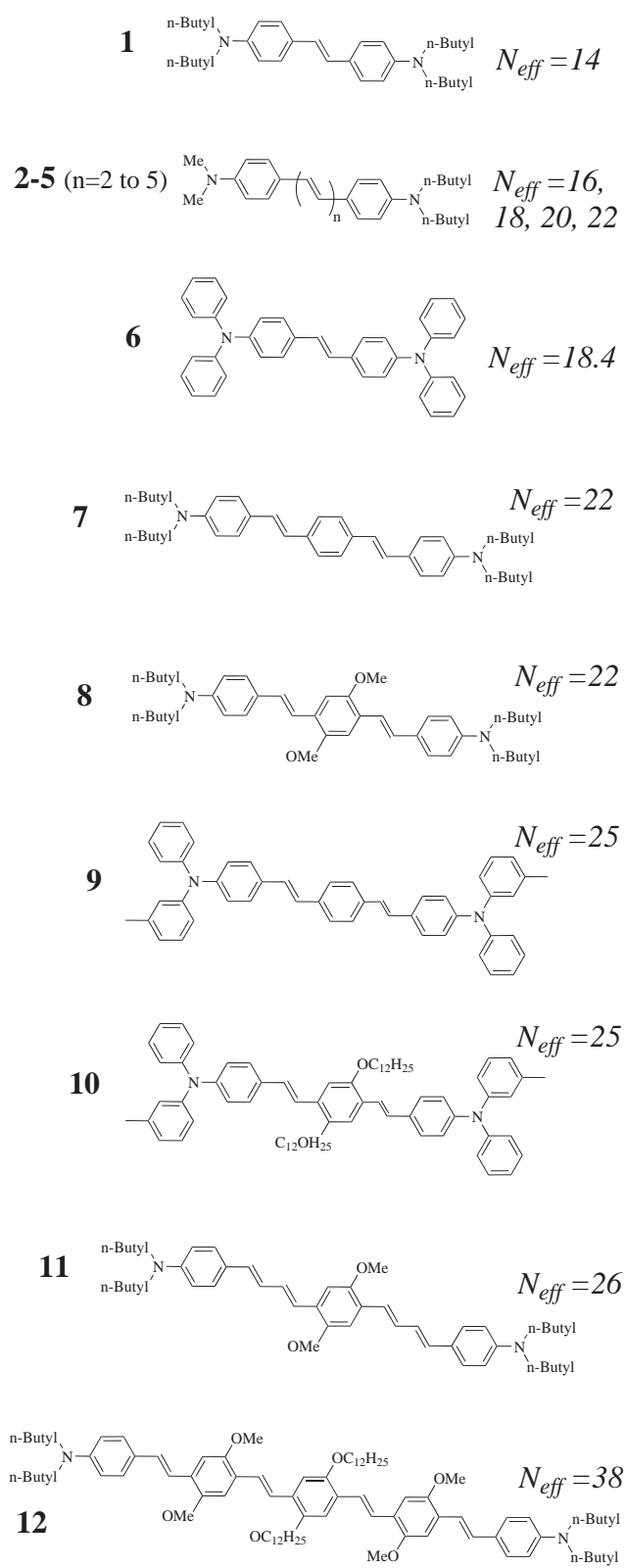} 
\caption{Molecules measured by Rumi and coworkers.\cite{Rumi}\label{fig:Rumi}}
\end{figure}

The molecules under analysis are listed in Figures \ref{fig:Rumi}, \ref{fig:albota}, and \ref{fig:Dhrovizhev}, together with the effective number of electrons, $N_{eff}$.  Note that the dendrimer structures, shown in Figure \ref{fig:Dhrovizhev}, from top to bottom shows molecules {\bf 20} and {\bf 21}.  The structures labelled G-1 and G-2 are the $R$-groups, that when added to molecule {\bf 21}, makes the dendrimers {\bf 22} and {\bf 23}.  The effective number of electrons are 18.4, 35.6, 55.8, and 82.7, respectively.

The TPA cross-section values of molecules $1$ to $12$ have been reported by Rumi and coworkers;\cite{Rumi} the values for molecules $13$ to $19$ have been reported by Albota and coworkers;\cite{Albota} and the values for molecules $20$ to $23$ have been reported by Drobizhev and coworkers.\cite{Drobizhev}  The value of $E_{10}$ for each molecule is calculated from the wavelength of maximum linear absorption. The values of the inverse radiative lifetimes are estimated to be: $\Gamma_{10} \approx \Gamma_{20} \approx 0.1 eV$.

We note that Frank-Condon considerations suggest that a more appropriate estimate for the energy of an excited state is the energy at the onset of absorption rather than the energy at the absorption peak.  For a typical excited state energy of $2 \, eV$ and width $100 \, meV$, the onset is about $50 \, meV$ from the peak, leading to a fractional error of $2.5 \, \%$.  For functions that vary as the fifth power of the energy, this contributes an error of less than $15 \, \%$ -- the low end for the experimental uncertainty of the more accurate nonlinear-optical measurements.  So, in using the peak values to obtain the energies, our calculations have an uncertainty of about $15 \, \%$; and, the positions of the peaks in our dispersion plots could be off by about $50 \, meV$.

\subsection{Influence of the number of electrons}

A simple consequence of the Thomas-Khun sum-rules is that the two-photon cross-section of a molecule depends quadratically on the effective number of electrons. With the aid of the sum-rules one can rewrite $\gamma_{I}^{TK3}$ as:
\begin{equation}
\label{eq:phidef}
\gamma_{I}^{TK\infty} = \left( \frac{\hbar e}{\sqrt{2m}} \right)^{2} \frac{N_{eff}^{2}}{E_{10}^{5}} \Phi,
\end{equation}
where $\Phi$ is a dimensionless function that depends on the energy ratios, the transition dipole moment ratios and the inverse radiative lifetimes but {\em does not depend on} $N_{eff}$. This is a general result that applies to molecules with any number of levels.  Eq. \ref{eq:phidef} together with Eq. \ref{eq:tpacross} predict a quadratic dependence on the effective number of electrons for the TPA cross-section:
\begin{equation}
\delta(\omega) \propto \frac{E_{20}^{2}}{E_{10}^{5}}N_{eff}^{2} \Phi.
\end{equation}

\begin{figure}
\includegraphics{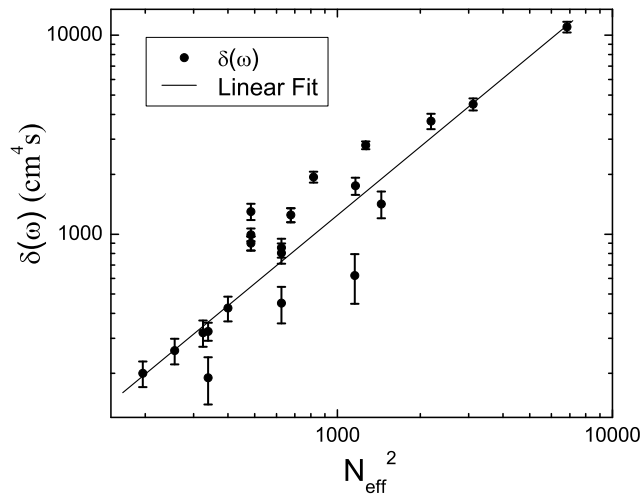} 
\caption{$\delta(\omega)$ as a function of $N_{eff}^{2}$ for the collection of measured molecules. The data fits to a line with correlation coefficient of $0.90394$.\label{fig:n2}}
\end{figure}

Fig. \ref{fig:n2} shows a plot of the TPA cross-sections of the different molecules as a function of $N_{eff}^{2}$.  The data follow an approximately linear relationship, indicating that even though the performance of a molecule is dictated by many factors, it correlates with the square of the effective electrons.  Note that while this scaling appears to be approximately universal for the molecules studied, it may not hold in other classes of molecules.

\subsection{Predicting the limiting values for real molecules}

Using a three-level model, from the experimental values of $N_{eff}$, $E_{10}$, $\Gamma_{10}$ and $\Gamma_{20}$, we can find the maximum value of $\gamma_{I}^{TK3}$ allowed by the sum rules, as follows.  From the theoretical section, the fundamental limit of $\gamma_{I}^{TK3}$ occurs when $X=1$, so the problem is reduced to finding the optimal value of $E_{20}$ that is consistent with the sum-rules.  We will denote this optimal value as $E_{20}^{max}$.

\begin{figure}
\includegraphics{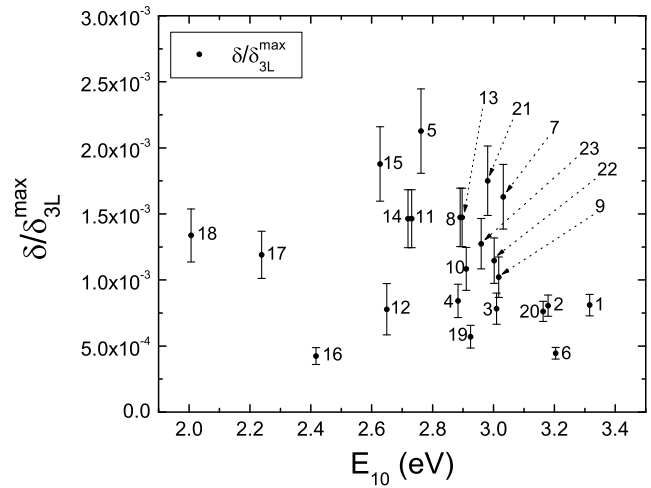} 
\caption{TPA performance as a function of $E_{10}$ for the collection of molecules studied here. The performance is evaluated by calculating the ratio between the experimental value of $\delta(\omega)$ and $\delta_{3L}^{max}$ (maximum value allowed by quantum mechanics in a three-level system).\label{fig:delta2norm}}
\end{figure}

The value of $E_{20}^{max}$ is substituted into Eq. \ref{eq:tpacross} to obtain the maximum value of $\delta(\omega)$, which we will denote $\delta_{3L}^{max}$.  Since in the resonant TPA regime, $E_{20}=2\hbar \omega$, $\delta_{3L}^{max}$ can be written as:
\begin{equation}
\label{eq:d3max}
\delta_{3L}^{max}=\frac{\pi (E_{20}^{max})^{2}}{\hbar n^{2} c^{2}} \langle {\gamma_{I}^{max}}^{*} \rangle,
\end{equation}
where $\langle {\gamma_{I}^{max}}^{*} \rangle$ is the orientational average of the fundamental limit of the dressed second hyperpolarizability.  In this manner, the maximum values of the TPA cross-section are calculated for the different molecules.  Since most of the molecules are geometrically one-dimensional, an isotropic orientational average yields $\langle \gamma_{I} \rangle = \gamma_{I}/5$.  The local fields are calculated using the Lorentz-Lorenz\cite{Jackson} model with $n=1.4$.   Fig. \ref{fig:delta2norm} shows the two photon performance (the ratio between the experimental value of $\delta(\omega)$ and $\delta_{3L}^{max}$).  

All the measured TPA cross-sections fall below the predicted limits by an average factor of $\approx 1.5 \times 10^{-3}$. This shows that real molecules are far below the maximum limit that is allowed by quantum mechanics.  Perhaps, as in the case of $\gamma_{I}$, the low performance ratio is due to the existence of many states beyond the dominant three-levels, which results in a dilution of the oscillator strengths.

The molecule with highest experimental TPA cross-section is {\bf 23} with $\delta(\omega)=(11000 \pm 1100) cm^{4}s$ - a similar performance to the rest of the molecules, even though its experimental value of $\delta(\omega)$ is 2 orders of magnitude higher than any other molecule in the collection. This indicates molecule {\bf 23} owes its high TPA cross-section mostly to the huge number of electrons contributing to the nonlinear response, rather than to a better design strategy that increases the effectiveness of those electrons. In fact, in the collection of molecules reported by Drobizhev and coworkers, the best performance is achieved by molecule {\bf 21}. Molecule {\bf 23} would need to have a $\delta(\omega)$ value of approximately $1.4$ times bigger than its experimentally-determined value in order to be as efficient as molecule {\bf 21}.

The best performance is achieved by molecule {\bf 5}. This is a somewhat odd result, because molecule {\bf 5} belongs to the group of homologues {\bf 1-5} and the performance of all the other homologues is much worse.  Rumi and coworkers report that the experimental uncertainty on molecule {\bf 5} is bigger than for the rest of their reported molecules, which could explain the discrepancy. The rest of homologues {\bf 1-4} behave as expected: the values of $\delta(\omega)$ increase as the number of electrons is increased, but the TPA performance is the same within experimental uncertainty and low in comparison to the other molecules.

This may be better understood if both excited state energies are considered.  First, the longer the homologue, the smaller the value of $E_{10}$ (red shifting), which increases the molecular performance due to the $1/E_{10}^{5}$ dependence.  However, for molecules {\bf 1} to {\bf 4}, there is also a red shifting of $E_{20}$ in such a manner that $E$ increases with the length of the homologue. The values of $E$ for molecules ({\bf 1}, {\bf 2}, {\bf 3}, and {\bf 4}) are respectively (0.802139, 0.820513, 0.86165, and 0.86165).  This increase of $E$ results in a lower performance.  However, although $E_{10}$ is red shifted for molecule {\bf 5} in comparison with the homologues, $E_{20}$ stays the same as for molecule {\bf 4}, resulting in a better-optimized energy ratio ($E=0.812918$).  Therefore, for molecules {\bf 1} to {\bf 4}, the red shifting of $E_{10}$ is accompanied by the red shifting of $E_{20}$, and the performance doesn't change appreciably.  In contrast, for molecule {\bf 5}, the increase in $1/E_{10}^{5}$ is not counterbalanced by an increase in $E$, so the performance of the molecule is increased.

Molecule {\bf 11} which extends the conjugated path of molecule {\bf 8} shows the same performance factor as its homologue and molecule {\bf 14} which extends the conjugated path of molecule {\bf 13} shows a performance factor that is the same as its homologue, within experimental uncertainty.

Molecules {\bf 8} and {\bf 13} have the same performance factor and also approximately the same value of $E_{10}$. Since they differ structurally only by the type of radicals attached to the central ring, we conclude that the effects of both types of radicals are similar in both the TPA performance and the wavelength of maximum absorption.  Similarly, molecules {\bf 11} and {\bf 14} have the same performance and similar values of $E_{10}$. Again, they only differ structurally by the type of radicals attached to the central ring; so we conclude that the effects of both types of radicals are similar.

Molecules {\bf 9}, {\bf 10}, {\bf 15} and {\bf 19} share a similar design strategy and differ by the type of groups attached to the central rings. This difference has an influence on the TPA performance. The best approach appears to be the attachment of CN radicals, which makes molecule {\bf 15} the second best molecule in the whole collection.

\subsection{Comparison with other models}

In the previous section, $\delta_{3L}^{max}$ was calculated by finding the optimal value of $E=E_{10}/E_{20}$ and taking the limit $X \rightarrow 1$. This requires ``tuning'' both the values of $E_{20}$ and $|\mu_{10}|$ for the given molecule.  A more realistic approach is to take the actual experimental values of $E_{10}$ and $E_{20}$, calculate $E$, substitute these values into $\delta(\omega)$ and calculate the maximum value possible for those energy values in the limit $X \rightarrow 1$. We will denote the maximum value calculated in this way as $\delta(E)^{max}$.

\begin{figure}
\includegraphics{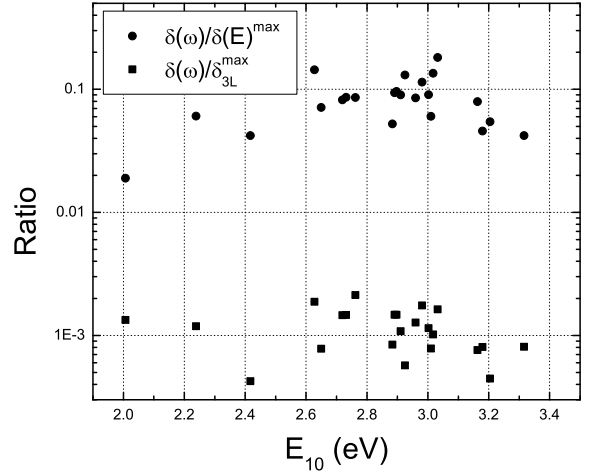} 
\caption{The ratio between the experimental value of $\delta(\omega)$ and $\delta(E)^{max}$ (maximum TPA cross-section allowed for the measured molecular energy ratios). This is compared with the ratio between $\delta(\omega)$ and $\delta_{3L}^{max}$ (maximum TPA cross-section allowed for that value of $E_{10}$).\label{fig:two_ratios}}
\end{figure}

Thus, for every molecule, we first compute $E$ as the ratio between the experimental values of $E_{10}$ and $E_{20}$ and then we calculate the maximum TPA cross-section allowed for that energy ratio, $\delta(E)^{max}$.  This can be compared with the maximum TPA cross-section possible for the same value of $E_{10}$ (with an optimized value of $E$), $\delta_{3L}^{max}$. Figure \ref{fig:two_ratios} plots the ratios of the experimentally-determined values normalized to each calculated limiting value, $\delta(\omega)/\delta_{3L}^{max}$ and $\delta(\omega)/\delta(E)^{max}$, as a function of $E_{10}$ for each molecule.

From Fig. \ref{fig:two_ratios} it is clear that the energy ratio, $E$, is far from optimized for all the molecules studied.  If it were possible to independently ``tune'' the value of $E_{20}$ and the value of $X$, the TPA cross-sections could be improved by three orders of magnitude. However, if the experimental value of $E_{20}$ stays fixed, the possible improvement for typical molecules such as the ones studies here (by tuning $X$) is only an order of magnitude.

To determine the relevance of the non-explicitly-resonant terms, the maximum TPA cross-section allowed for a particular value of $E$ for a molecule, $\delta(E)^{max}$, can be compared with the maximum value that would be possible if the response were determined only by the explicitly resonant contribution, $\gamma_{res}$ (Eq. \ref{resonant}). We will denote this maximum value $\delta_{res}$.  

We can also calculate the maximum values predicted by Kuzyk's approximation technique of extrapolating from the off-resonance results.\cite{Kuztwopho}  This off-resonance extrapolation technique (ORET) assumes that $\gamma_{I}^{TK3}$ is dominated by $\gamma_{res}$, and predicts the following:
\begin{eqnarray}\label{eq:ORET-RES}
\delta_{ORET}^{res} & = & 63.5 \left\{ \frac{1}{n^{2}} \left( \frac{n^{2}+2}{3} \right)^{4} \right\} \\ \nonumber
& \times & \left[ \frac {\left( E_{10} - \frac{E_{20}} {2} \right)^2 - \Gamma_{10}^2} {\left( \left( E_{10} - \frac{E_{20}} {2} \right)^2 + \Gamma_{10}^2 \right)^2 } \right] \left( \frac{E_{20}}{\Gamma_{20}} \right) \left( \frac{N_{eff}^{2}}{E_{10}^{3}} \right).
\end{eqnarray}
In the off-resonance case, when $\Gamma_{n0} \ll E_{10} - E_{20}/2$, Eq.~\ref{eq:ORET-RES} can be approximated by:
\begin{eqnarray}\label{eq:ORET-offRES}
\delta_{ORET} & = & 63.5 \left\{ \frac{1}{n^{2}} \left( \frac{n^{2}+2}{3} \right)^{4} \right\} \\ \nonumber
& \times & \left[ \frac {1} { E_{10} - \frac{E_{20}} {2} } \right] \left( \frac{E_{20}}{\Gamma_{20}} \right) \left( \frac{N_{eff}^{2}}{E_{10}^{3}} \right).
\end{eqnarray}

\begin{figure}
\includegraphics{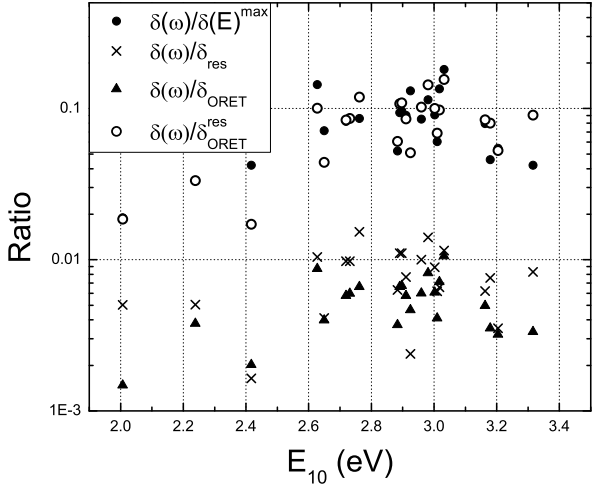} 
\caption{The ratio between the experimental value of $\delta(\omega)$ and $\delta(E)^{max}$, $\delta_{res}$, $\delta_{ORET}$ and $\delta_{ORET}^{res}$.\label{fig:three_ratios}}
\end{figure}

Fig. \ref{fig:three_ratios} plots the ratio between the measured TPA cross-section, $\delta(\omega)$, and the various calculated limits: $\delta(\omega)/\delta_{res}$, $\delta(\omega)/\delta_{ORET}$, and $\delta (\omega) / \delta (E)^{max}$, and $\delta (\omega) / \delta_{ORET}^{res}$, which shows that both $\delta_{ORET}$ and $\delta_{res}$ are about an order of magnitude higher than $\delta (E)^{max}$.   (Recall that $\delta (E)^{max}$ is the fundamental limit of TPA absorption using all terms in the three-level model for a fixed $E$ that is determined from experiment.)  The ORET extrapolation using the resonant expression (Eq.~\ref{eq:ORET-RES}) appears to be of the same order of magnitude as the exact calculation while the off-resonant approximation (Eq. \ref{eq:ORET-offRES}) deviates by about an order of magnitude.  Table \ref{tab:exact_hypothesis} shows a comparison of the ORET extrapolation and the exact results.  On average, variations between the ORET extrapolation and the exact results are just over 30\% -- the variation being the price of ignoring the non-explicitly resonant contributions to $\gamma_{I}^{TK3}$.
\begin{table}
\caption{The values of $\delta(E)^{max}$ compared with the results of Eq.~\ref{eq:ORET-RES} for the set of molecules studied here, and the percentage difference between the exact result and the the resonant ORET formula (Eq.~\ref{eq:ORET-RES}).\label{tab:exact_hypothesis} }
\begin{eqnarray*}
\begin{array}{l l l l } \hline
& &  \\ 
\mbox{Molecule} & \delta(E)^{max} & \delta_{ORET} & \%   \\
& \mbox{(GM)} & \mbox{(GM)} & \mbox{ Diff.} \\ 
& &  & \\ \hline \hline
1 & 4745 &    2213  & 53 \\ 
2 & 5664 &    3250  & 43 \\ 
3 & 5291 &    4663  & 12 \\ 
4 & 8110 &    7001  & 14 \\ \hline
5 & 15124 &   10936 & 28 \\
6 & 3482 &    3598  & 3  \\ 
7 & 5480 &    6395  & 17 \\ 
8 & 9574 &    8358  & 13 \\ \hline
9 & 5960 &    8232  & 38 \\ 
10 & 9474 &   10019 & 6 \\  
11 & 14515 &  14523 & 0.06 \\ 
12 & 19931 &  32272 & 62 \\ \hline
13 & 9338 &   8247  & 12 \\
14 & 15212 &  14894 & 2 \\ 
15 & 13456 &  19322 & 44 \\ 
16 & 14695 &  36080 & 146 \\ \hline
17 & 28897 &  52513 & 82 \\
18 & 195446 & 199497 & 2 \\
19 & 3432 &   8835   & 157 \\ 
20 & 4090 &   3862   & 6 \\ \hline
21 & 24460 &  19519  & 20 \\
22 & 49604 &  44996  & 9 \\
23 & 129198 & 107574 & 17 \\ \hline \hline
\end{array}
\end{eqnarray*}
\end{table}

Next, we consider extrapolating the off-resonance ORET term, given by Eq.~\ref{eq:ORET-offRES}.  So, while the $\delta_{ORET}$ values are consistently larger than $\delta (E)^{max}$ for each molecule, they are approximately related logarithmically by:
\begin{equation}
\ln (\delta(E)^{max}) =  \lambda \ln(\delta_{ORET}),
\end{equation}
or
\begin{equation}
\label{eq:lambda}
\delta(E)^{max}=(\delta_{ORET})^{\lambda},
\end{equation}
where $\lambda \approx 0.773$.

To illustrate the validity of Eq. \ref{eq:lambda}, we list the values of $\delta(E)^{max}$ and $(\delta_{ORET})^{0.773}$ for the different molecules in Table \ref{table:lambda-hypothesis}.
Within $\pm 20 \%$, Eq.~\ref{eq:lambda} is a good approximation for most of the molecules. The case of molecule {\bf 19}, which has the largest variance, will be discussed later in terms of the consistency of the three-model. Thus, in our collection of 23 molecules there are only 4 outside the $\pm 20 \%$ range.  As such, Eq. \ref{eq:lambda} is a good approximation of the limit, which includes the effects of the contributions of the non-explicitly-resonant terms to $\gamma_{I}^{TK3}$.  This approximation is simpler to use and is more accurate than the resonant expression, so provides the researcher with a simple expression to evaluate molecules.
\begin{table}
\caption{The values of $\delta(E)^{max}$ and $(\delta_{ORET})^{0.773}$ (Eq. \ref{eq:lambda}) for the set of molecules studied here, and the percentage difference between the exact result and the approximation (Eq. \ref{eq:lambda}).\label{table:lambda-hypothesis}}
\begin{eqnarray*}
\begin{array}{l l l l } \hline
& &  \\ 
\mbox{Molecule} & \delta(E)^{max} & \delta_{ORET}^{0.773} & \%   \\
& \mbox{(GM)} & \mbox{(GM)} & \mbox{ Diff.} \\ 
& &  & \\ \hline \hline
1 & 4745 & 4932 & 4 \\ 
2 & 5664 & 5802 & 2 \\ 
3 & 5291 & 6054 & 14 \\ 
4 & 8110 & 8142 & 0.4 \\ \hline
5 & 15124 & 12337 & 18 \\
6 & 3482 & 4899 & 41  \\ 
7 & 5480 & 7002 & 28 \\ 
8 & 9574 & 9302 & 3 \\ \hline
9 & 5960 & 8050 & 35 \\ 
10 & 9474 & 9927 & 5 \\ 
11 & 14515 & 12958 & 11 \\ 
12 & 19931 & 19601 & 2 \\ \hline
13 & 9338 & 9171 & 2 \\ 
14 & 15212 & 13308 & 13 \\ 
15 & 13456 & 13621 & 1 \\ 
16 & 14695 & 17432 & 19 \\ \hline
17 & 28897 & 23953 & 17 \\
18 & 195446 & 88272 & 55 \\
19 & 3432 & 7155 & 108 \\ 
20 & 4090 & 5283 & 29 \\ \hline
21 & 24460 & 19015 & 22 \\ 
22 & 49604 & 34445 & 31 \\ 
23 & 129198 & 69466 & 46 \\ \hline \hline
\end{array}
\end{eqnarray*}
\end{table}

\subsection{Consistency of the three-level model}

In the previous section, $\delta(E)^{max}$ was calculated by substituting the experimental values of $E_{10}$ and $E_{20}$ into $\delta(\omega)$ and taking the limit $X \rightarrow 1$. Alternatively, with the experimental values of $E_{10}$, $E_{20}$ and $\delta(\omega)$ we can find the value of $X$ that, using the three-level model, would yield the same experimental value $\delta(\omega)$.

So, for a given molecule, we can calculate $X^{3L}$, the ratio of $|\mu_{10}|/|\mu_{10}^{max}|$ that is derived from the sum-rule-restricted sum-over-states three-level model and the experimental data.  Three different scenarios might result:
\begin{enumerate}
\item The value of $X^{3L}$ is physically impossible (i.e. an imaginary quantity, $X^{3L} >1$, etc). In this case we can conclude that the three-level model does not describe the molecule.

\item The value of $X^{3L}$ is physically allowed but inconsistent with the typical values of $X$ for organic molecules. For instance, the three-level model could predict $X^{3L}=10^{-4}$ which is possible but highly unlikely for real molecules.  In this case we would also conclude that the three-level model is not likely to accurately approximate the molecule's TPA cross-section. 

\item The value of $X^{3L}$ is consistent with physically-allowed values of $X$ in a range that is typical for organic molecules.  In this case, although we can not conclude definitively that the three-level model accurately describes the molecule, the reasonableness of the result suggests that it may be suitable as a model for real molecules.
\end{enumerate}

Table \ref{tab:X} lists the values of $X^{3L}$ for the collection of molecules reviewed here.  The predicted values are all consistent with what is normally observed for typical molecules, except for molecules {\bf 16} and {\bf 19}, which yield complex values of $X$.  Note that these two molecules show the largest discrepancies in Table \ref{tab:exact_hypothesis}.  Clearly, for these two molecules, the three level model does not do a good job of modelling the TPA cross-section.  Interestingly, these two molecules (excluding {\bf 6}) have the smallest TPA performance ratio, a result that is consistent with our assertion that molecules with more contributing excited states are less efficient per electron.  The rest, however, can consistently be described by our simplified three-level model.

\begin{table}
\caption{The values of $E$ and $X^{3L}$ for typical TPA molecules studied here. $E$ is calculated from the experimental values of $E_{10}$ and $E_{20}$. $X^{3L}$ is calculated by using it as a floating parameter when fitting the TPA cross-section data to the three-level model as restricted by the sum-rules.\label{tab:X}}
\begin{eqnarray*}
\begin{array}{ c c c } \hline
& &  \\ 
\mbox{Molecule} & \mbox{Energy Ratio} & X^{3L} \\
& E &  \\ \hline \hline
1 & 0.80214 & 0.26118 \\ 
2 & 0.82051 & 0.2633  \\ 
3 & 0.86165 & 0.26714 \\ 
4 & 0.84884 & 0.25194 \\ \hline
5 & 0.81292 & 0.36609 \\ 
6 & 0.89147 & 0.17847 \\ 
7 & 0.89242 & 0.47829 \\ 
8 & 0.85082 & 0.36386 \\ \hline
9 & 0.90633 & 0.36377 \\ 
10 & 0.87441 & 0.32751 \\ 
11 & 0.85352 & 0.33781 \\ 
12 & 0.89744 & 0.13118 \\ \hline
13 & 0.8528  & 0.36753 \\ 
14 & 0.84978 & 0.33168 \\ 
15 & 0.88453 & 0.4101 \\ 
16 & 0.91618 &  \\ \hline
17 & 0.87545 & 0.12717 \\ 
18 & 0.76456 & 0.15067 \\ 
19 & 0.9434 &  \\
20 & 0.8801 & 0.30138 \\ \hline
21 & 0.84736 & 0.40954 \\ 
22 & 0.86441 & 0.34658 \\ 
23 & 0.85084 & 0.34567 \\ \hline \hline
\end{array}
\end{eqnarray*}
\end{table}

\section{Conclusions}

We have applied the sum rules to the SOS expression to calculate the fundamental limits of the dispersion of the two-photon absorption cross-section.  These results apply at all wavelengths, so can be used both on and off resonance.  Our new rigorous analysis shows that the three-level model and truncated sum rules together give a reasonable fundamental limit and is a good approximation when modelling most molecules.  This conclusion partly follows from an analysis that applies the sum rules to the resonant terms with an infinite number of states, and shows that the resonant term is at its maximum in the three-level limit.  This result is significant because it suggests that our ansatz that the three-level model accurately describes molecules whose nonlinear susceptibilities approach the fundamental limit is reasonable.

We have shown that the non-explicitly-resonant terms, which are often ignored, can not be neglected in real molecules even near resonance since these terms have a profound effect on resonant dispersion.  As such, experiments measuring TPA cross-sections that are analyzed near resonance using only the resonant terms may yield unreliable conclusions.

We find that the simple process of extrapolating the fundamental limit of the off-resonant two-photon absorption cross-section to resonance using only the resonant term of the three-level model yields a result that correlates well with our exactly rigorous results.  A logarithmic relationship is proposed that can be used to convert the off-resonance extrapolation result to the exact one to within an accuracy of about $\pm 20 \%$.  This relationship is much easier to apply than the full rigorous result and should be useful to researchers who are interested in assessing the TPA performance ratio of a molecule.

The TPA cross-sections of existing molecules are found to be far from the fundamental limit; so, there is room for improvement.  To reach the fundamental limit would require precise control of the energy-level spacing, independently of the transition dipole moments --  a task that does not appear possible using today's synthetic approaches.  To make the task within reach, we present an analysis of how a molecule can be maximized for a given energy ratio, so, researchers only need to focus on adjusting the oscillator strength to the first excited state.  While this approach will not result in molecules with TPA cross-sections at the fundamental limit, it can yield a factor of ten improvements in the largest TPA cross-sections measured.

Clearly it is best to normalize TPA measurements to the fundamental limits when comparing molecules; but, we have shown that simply dividing by the square of the number of electrons per molecule yields a good metric for comparison.  Such a simple performance metric is most attractive for cases where the transition moments and energy levels of a molecule are unknown, which is often the case.

In summary, we report on a mix of theoretical results and analysis techniques that will both aide in building a deeper understanding of TPA cross-sections and provide the experimenter with quick methods for assessing measurements.

\section{Acknowledgements}

We thank Wright Paterson Air Force Base and the National Science Foundation (ECS-0354736) for generously supporting this work.  Javier Perez-Moreno acknowledges a PhD fellowship from the University of Leuven and support from the Department of Physics and Astronomy at Washington State University and the Department of Chemistry at the University of Leuven.


\end{document}